\documentclass[sigconf]{acmart}

\usepackage{algorithmic}
\usepackage{graphicx}
\usepackage{textcomp}
\usepackage{xcolor}
\usepackage{rotating}
\usepackage{multirow}
\usepackage{float}
\usepackage{listings}
\usepackage[english]{babel}
\usepackage{hvfloat}
\def\BibTeX{{\rm B\kern-.05em{\sc i\kern-.025em b}\kern-.08em
    T\kern-.1667em\lower.7ex\hbox{E}\kern-.125emX}}

\newcommand{\observe}[1]{
\noindent\fbox{%
    \parbox{.95\columnwidth}{%
        {#1}
    }%
}}

\setcopyright{acmcopyright}
\copyrightyear{2018}
\acmYear{2018}
\acmDOI{XXXXXXX.XXXXXXX}

\acmBooktitle{Woodstock '18: ACM Symposium on Neural Gaze Detection,
  June 03--05, 2018, Woodstock, NY} 
\acmPrice{15.00}
\acmISBN{978-1-4503-XXXX-X/18/06}

\theoremstyle{definition}
\newtheorem{definition}{Definition}[section]

\begin{document}

\title{Defining and executing temporal constraints for evaluating engineering artifact compliance}


\author{Cosmina-Cristina Ratiu}
\email{cosmina-cristina.ratiu@jku.at}
\author{Christoph Mayr-Dorn}
\email{christoph.mayr-dorn@jku.at}
\author{Alexander Egyed}
\email{alexander.egyed@jku.at}
\affiliation{%
  \institution{Johannes Kepler University, Institute for Software Systems Engineering}
  \city{Linz}
  \country{Austria}
}

\renewcommand{\shortauthors}{Ratiu et al.}

\begin{abstract}
Engineering processes for safety-critical systems describe the steps and sequence that guide engineers from refining user requirements into executable code, as well as producing the artifacts, traces, and evidence that the resulting system is of high quality. 
Process compliance focuses on ensuring that the actual engineering work is followed as closely as possible to the described engineering processes. To this end, temporal constraints describe the ideal sequence of steps. Checking these process constraints, however, is still a daunting task that requires a lot of manual work and delivers feedback to engineers only late in the process.
In this paper, we present an automated constraint checking approach that can incrementally check temporal constraints across inter-related engineering artifacts upon every artifact change thereby enabling timely feedback to engineers on process deviations. Temporal constraints are expressed in the Object Constraint Language (OCL) extended with operators from Linear Temporal Logic (LTL).
We demonstrate the ability of our approach to support a wide range of higher level temporal patterns. We further show that for constraints in an industry-derived use case, the average evaluation time for a single constraint takes around 0.2 milliseconds.
\end{abstract}



\maketitle

\section{Introduction}

The development process of safety-critical systems must adhere to several safety regulations and standards, such as ASPICE or ISO 26262 in the automotive industry \cite{oliveira2017analysis}. 
The standards describe various methods for software developing organizations to apply in their engineering processes to ensure high quality software \cite{damian2006empirical}. The standards typically also mandate evidence of how those methods are applied and checked.
The resulting processes that describe how the system, respectively, the software, is to be developed serve primarily as reference documentation for engineers to check their activities against. The processes can never be rigorously enforced as engineering is inherently an iterative work that requires a significant amount of flexibility for engineers to handle unanticipated new information, changing customer feedback, corner cases, ambiguities, etc.
Consequently engineers, and most often, quality assurance personnel manually check engineering artifacts (e.g., requirements, specification, test cases, reviews, source code), traces among these artifacts, and their timestamps to understand to what extent the actual engineering process deviates from the intended one.
From a process perspective, deviations typically come in the form of incorrect order of changes, e.g., design specification artifacts are changed first, and corresponding requirements are only updated later; changes to user requirements are implemented without first going through a review, with a review only happening much later. These deviations then result in lower quality engineering artifacts, rework, and delays.


Detecting deviation in a timely manner is a challenge, as most of the process-centric constraints 
are checked manually. This procedure typically requires looking for any inconsistencies between artifact change timestamps and the expected sequence of the engineer process steps. This analysis is highly time-consuming and repetitive, thus typically only conducted at particular miles stones (e.g., upon requirement refinement completion) in the engineering process when artifacts are completed and ready to be used for a subsequent stage. The major downside of this infrequent checking is that QA personnel may have to request clarifications on change sequences (e.g., why a requirement change has not led to a review) from engineers when they have already moved on to another task and have to recall that situation. Also, any process deviation feedback is communicated to developers long after the fact, which reduces the opportunities to correct earlier (where possible) and remain more compliant with the process.
One way to mitigate this problem is through automatic checking of temporal constraints.

Existing approaches fall into three broad categories: first, work on artifact constraint checking (i.e., consistency checking)  \cite{Egyed11TSE,konig2016advanced, 10.1145/3270112.3275335}---while focusing on engineering artifacts---cannot express temporal constraints. Second, work on temporal constraint checking at design time (e.g., using formalisms such as Linear Temporal Logic - LTL) focuses on ensuring that a process, model, or specification will not lead to undesirable (i.e., deviating) behavior but are not applicable as the engineering process cannot force engineers to stay within the process definition. Third, work on run-time verification compares a sequence of events against a formula (again often using LTL) to detect deviations. Few of those latter approaches support data properties in their constraint definition, and none of them enable the definition of temporal constraints over arbitrary collections of and relations between artifacts. This aspect is crucial in engineering processes that largely manifest as changes to interrelated artifacts, for example, to specify that all accepted user requirements that are linked to a particular change request and are assigned to the latest release must go through a review upon every update.

Hence, in this paper, we are introducing a mechanism for data-centric temporal constraint checking. To this end, we extend the well-known object constraint language (OCL) \cite{ali2014insights} with temporal operators from LTL as the foundation for expressing temporal constraints over arbitrary artifact relations. 
The primary novelty is in our mechanism, which then continuously evaluates these temporal constraints upon every artifact change to provide continuous, timely feedback to engineers upon constraint fulfillment, respectively violation.

We evaluate our proposed mechanism along two lines: we are investigating its  expressiveness and performance based on a process excerpt from our industry partner ACME-RA. We evaluate expressiveness by demonstrating that we can support all state-of-the-art temporal constraint specification patterns proposed in DECLARE~\cite{declare,de2014reasoning} (which in turn are an extension to the property specification patterns of Dwyer et al.~\cite{dwyer1998property}). 
For evaluating performance, we collected a set of temporal constraints based on the development process from our industry partner ACME-RA, subsequently replaying over 57,000 artifact changes in the scope of 267 engineering process instances to determine whether our constraint checking mechanism is rapid enough to make timely feedback possible. Our results show that for this realistic case study, our mechanism is able to evaluate a single temporal constraint on average in $\sim$0.2 milliseconds.

Our salient contributions in this paper are:
\begin{itemize}
    \item an extension to OCL for specifying temporal logic based on LTL operators over arbitrary artifact and their relations.
    \item a prototype for continuously evaluating temporal constraints.
\end{itemize}

In the remained of this paper, we are first introducing a motivational example based on the experience of our industry partners in Section~\ref{sec:motivatingscenario}, followed by a discussion of related work in Section~\ref{sec:sota}. We then provide the details of our novel temporal constraint checking mechanism in Section~\ref{approach}. Subsequently, we describe our evaluation design, results, and their implications in Section~\ref{evaluation}.
We conclude this paper with a summary and outlook on future work in Section~\ref{conclusion}.

\section{Motivating Scenario} \label{sec:motivatingscenario}

In the following, we will introduce an illustrative scenario that is based on several discussions with our industry partner ACME-AUTO from the automotive industry. The systems developed in this sphere have to comply with strict safety regulations defined in the ASPICE standard, and the engineering process is organized based on the V-model.

\begin{figure}[!htbp]
\caption{\label{fig:processexcerpt} Motivating engineering process excerpt.}
\centering
\includegraphics[width=0.8\columnwidth]{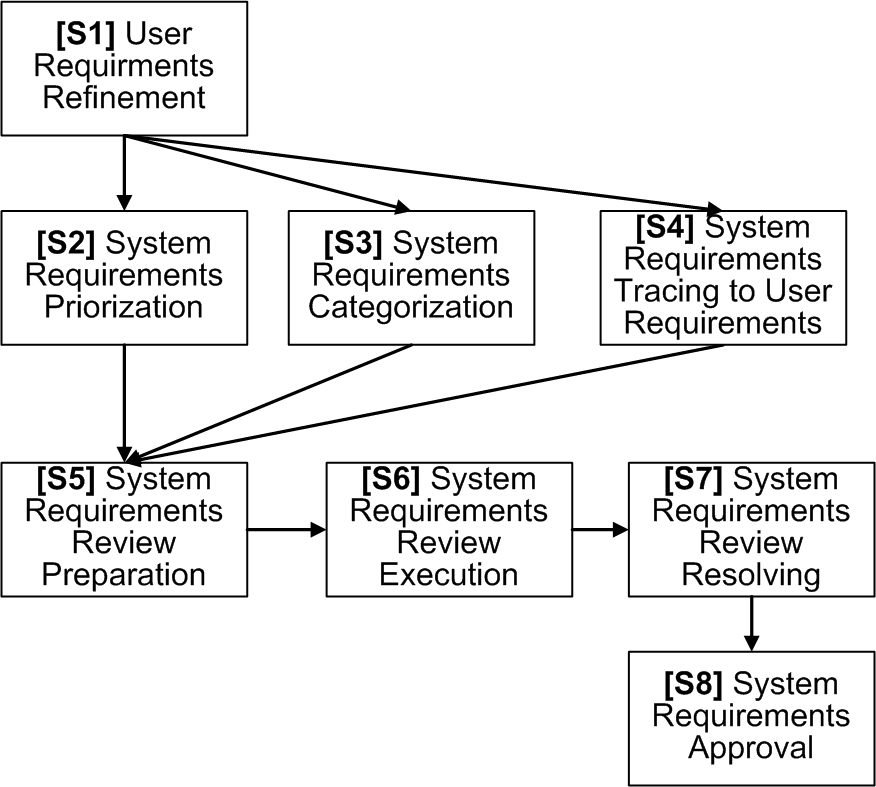}
\end{figure} 

Figure \ref{fig:processexcerpt} depicts an excerpt of an exemplary, simplified engineering process describing the refinement of user requirements to system requirements. The involved steps comprise the actual requirement refinement, subsequent prioritization, categorization, and tracing before reviewing the requirements for understandability, feasibility, consistency, and completeness. Ultimately system requirements are approved for subsequent design and implementation phases.
Typically engineers signal the progress on individual system requirements via the requirement's status, such as ``Draft'', ``Ready for Review'',  or ``Reviewed''.  These states support engineers in understanding when their work can start and guides them on how to signal that they consider their work as done. For example, the reviewing step should only start when \textbf{all} system requirements associated with a change request are in state ``Ready for Review''.

Without any automated constraint checking support, engineers need to manually check when this condition is fulfilled, or no longer fulfilled. Moreover, engineers need to ensure that whenever a requirement is updated that it then goes through the reviewing step again. Often dedicated Quality Assurance engineers check that this sequence is followed. Given the hundreds of artifacts to check across multiple projects and constraint variations, feedback often arrives late - hence the need for automated checking.
Existing constraint checking approaches \cite{Egyed11TSE,konig2016advanced, 10.1145/3270112.3275335} are not applicable for this purpose as these focus on conditions that need to remain fulfilled at all points in time. Artifact states, however, change over time. Our example constraint is typically fulfilled just before a review step, but will become violated as soon as the first system requirement is marked as ``Reviewed''. Ultimately, we need to be able to express that all system requirements are in state ''Ready for Review'' at some stage of their life-cycle, and, additionally, that once a system requirement is in status ``Draft'' (again) that it also goes into status ``Ready for Review'' thereafter. We might even want to additionally ensure that all the system requirements are in that status at the same time.

The key challenge that motivates this paper is the need to check temporal constraints across multiple related artifacts. In other words, it is insufficient to be able to check the correct sequence of only an individual artifact's property changes. Simply ensuring, for example, that a single requirement's status changes adhere to a temporal constraint is too limiting. In our motivating scenario, a constraint checking mechanism needs to be able to navigate from a change request to all associated (e.g., linked) system requirements and access their status property and evaluate changes to the status property. It also needs to handle the situation when the list of linked requirements grows as the engineers refine the individual user requirements into system requirements. An automated temporal-aware constraint checking mechanism can provide immediate feedback to engineering upon every small change to any engineering artifact while a quality assurance engineer doesn't have the capacity to accomplish this in the same timely manner.

\section{Related Work}
\label{sec:sota}
We start our discussion of related approaches according to their expressiveness
followed by an analysis of design time and run time checking strategies used.

\subsection{Constraint Specification Expressiveness}

The constraint specification literature holds a wide variety of approaches focused both on design time and run time compliance checking. 
Metric first-order temporal logic (MFOTL)~\cite{basin2010policy} exhibits very similar expressiveness in terms of describing temporal constraints over artifact relations but takes a very different approach to encoding those constraints. MFOTL builds on relational algebra, and hence dedicated predicates, to express relations among artifacts. For example, the MFOTL equivalent of OCL's ``requirement.changeRequest'' is expressed as a ``hasCR(req,cr)'' predicate. There are some expressiveness limitations, however, such as the inability to access elements from a list (which OCL is capable of) or OCL functions such as ``String.contains()''. Formulae containing relations/functions such as the latter that are domain-dependent are not supported as they cannot be inverted without resulting in infinite relations (which is required by MFOTL to evaluate such formulae).
Another downside is the difficulty of expressing non-trivial constraints that involve multiple relations among artifacts. Authors of OCLR and TempSy have reported that constraints in MFOTL were difficult to formulate even for an expert having ``ten years of experience in formal specification (and verification) with temporal logic''~\cite{dou2017tempsy}.
In practice, the use of relational algebra, Petri Nets, or similar rigorous specification languages can easily become prohibitive to software engineers and process modelers with limited experience in the model-checking field~\cite{hsueh2008applying}. 

Alternatively, Temporal OCL~\cite{kanso2013temporal}, OCLR~\cite{dou2014oclr,dou2017model,dou2017tempsy} and similar approaches~\cite{al2013approach,6958405,soden2009temporal,ziemann2003ocl} extend OCL to describe constraints on the behavior of a system in terms of method call sequences. They have in common to build on Dwyer et al.'s patterns~\cite{dwyer1998property} rather than LTL.
They also share a common limitation: Temporal OCL, OCLR, and Tempsy introduce a new dedicated OCL root element containing a set of patterns over temporal events. An event can be the fulfillment of any traditional OCL expression or the calling of a method (with specific pre and post conditions). Nesting of temporal constraints inside the traditional OCL expressions, however, is not possible. 

\subsection{Constraint Checking Strategy}

\subsubsection{Process Checking at Run Time}

When a process cannot be defined, or participants cannot be controlled tightly enough, run-time checking detects process deviations and provides feedback. While most of these approaches can also be applied at design time, they offer additional support for process monitoring during execution. They also often support varying degrees of process deviation, allowing the users to violate certain constraints, either temporarily while being offered some repair suggestions~\cite{knuplesch2017framework}, or permanently for a selection of ``optional constraints''~\cite{pesic2007constraint}.

In most cases, run time checking is achieved through complex event processing techniques~\cite{walzer2007temporal,walzer2008relative,cugola2012processing, montali2014monitoring}. These techniques observe the flow of atomic events generated by changes during the process. Based on this inflow, the constraints check for temporal patterns~\cite{awad2015runtime, barnawi2015bp} that define violations of the regulations, such as the order of the events being incorrect. Alternatively, some approaches use methods similar to those in process design time checking, such as translating the constraints into automata\cite{maggi2011monitoring} or ontologies\cite{elgammal2015lifecycle} before checking them over the stream of events.

All these approaches have in common 
that they work with events that do not capture the context of a change in terms of the structure and relations of the affected artifacts. However, in engineering processes, a lot of changes affect interrelated artifacts. As a result, contemporary complex event processing techniques do not support the navigation across data (i.e., artifacts) needed to express the temporal constraints relevant in engineering environments. 

\subsubsection{System Checking at Run Time}

The run time verification of system models can be performed during the execution of the system, or after the fact, based on the trace of changes collected during the execution.

For just-in-time checking, one approach by Bill et al.\cite{bill2014model} is to generate all the possible future states of the system based on the system model and the current state it is in, and check if any of these future possible states violates the temporal constraints specified. This approach shares the downsides of design time checking. 

Other approaches capture the state of the system at different time points during the execution of the process, collecting them to form the trace of the system's execution history. Based on this trace, they are validating temporal constraints regarding the order and timing of the changes~\cite{dou2014oclr,dou2017tempsy,dou2017model}, similarly to complex event processing approaches for run time process checking. As a result, they also lack contextual information regarding the relation of each event to the structure of the overall system. 

Alternatively, approaches such as that proposed by Basin et al.,~\cite{basin2010policy,basin2017monpoly} express constraints as sets of predicates, the evaluation of which is stored in a log file with timestamps capturing the temporal aspect of the execution. Then, this resulting series of predicate evaluations can be used to check temporal constraints through specialized model checking tools.

\subsubsection{Design Time Checking}

The approaches in this category focus on checking the compliance of process and system models to regulations at design time. Namely, they analyze these models and determine whether there are paths or possible future states that violate specific constraints, such as pre and post conditions that conflict, or processes that can never terminate.

To this end, a common approach is generating all the possible paths through the model~\cite{laurent2014alloy4spv,holfter2019checking,haarmann2019compliance,mendoza2014business,7723626,al2013approach,6958405}. Then, the temporal constraints specified based on the set of compliance regulations, are applied on these paths~\cite{schumm2010business,becker2011modeling,soden2009temporal,ziemann2003ocl,humberg2013ontology,pham2016checking,elgammal2015lifecycle}. The aim is to identify all the  paths that lead to constraint violations or, alternatively, to prove that no such path exists. In system models defined using UML, different OCL extensions, such as that proposed by Kanso et al.\cite{kanso2013temporal} patterns to specify temporal constraints compatible with OCL constraint specifications. 

This is useful for process model validation, i.e., ensuring that the process or system model is consistent and will not result in an undesirable execution sequence.
In engineering environments, however, the engineers cannot be restricted to such an extent that any deviation is avoided altogether.

\section{Approach}
\label{approach}
In this section, we describe our novel temporal constraint checking approach that exhibits three key distinguishing features: (1) evaluation occurs incrementally, (2) temporal OCL constraints may be defined across arbitrary artifact relations (e.g., traces), and (3) temporal operators may occur at any location in an OCL constraint.

As seen in Section~\ref{sec:sota}, in practice there are two common interpretations (respectively implementations) of applying temporal logic. On the one hand, approaches like Alloy4SPV\cite{laurent2014alloy4spv} check that constraints over a system's state or properties hold over a \textbf{continuous} period of time. On the other hand, process-centric checking approaches such as DECLARE\cite{declare}---or complex event process approach in general---are applied over traces of \textbf{discrete} events (e.g., starting or finishing of process steps) to evaluate whether the order of these events complies with the temporal constraints.

Our approach falls into the first category. Temporal patterns \cite{6227125}, however, are typically defined with discrete events in mind. Hence, as part of the evaluation of our approach's expressiveness and applicability, we provide a mapping of the comprehensive set of higher level patterns in DECLARE to our approach and their resulting interpretation. Note that in both categories, discrete events occur. In the former, events are changes to the system's (or model's) state which trigger evaluation of whether affected constraints (still) hold, while in the latter category, events are directly subject to the temporal constraints. 

In this section, we start first by defining what time and events mean in our approach, followed by an elaboration on the traditional LTL operators.



\subsection{Fundamentals}

In engineering processes, the subject of interest is the engineering artifacts and their relations, such as various types of requirements, specifications, test cases, source code, etc. but also coordination-centric engineering artifacts, such as change requests, stories, epics, bugs, work packages, and so on. We treat all these as a connected graph of artifacts across which we navigate as needed. 

Each constraint defined in the artifact graph has a specified context (e.g., a change request). Rule evaluation then accesses some properties of this context, potentially navigating to addition artifact and their properties, thereby forming the scope of the constraint. As the model changes, a constraint's scope can change as well. For example, new artifacts are added to or removed from collection properties (e.g., adding a requirement to a change request's list of related requirements). We define the scope of a constraint as follows.

\begin{definition}[Constraint Scope]
\label{def:scope}
    Let \(A\) be the set of artifacts in the model, and for each $a \in A$, \(P_a\) the set of properties of artifact a. Let \(C\) be the set of constraints specified in the model. The tuple (\(a\),\(p_a\))  is \textit{in the scope} of constraint $c \in C$ if 
    checks the value of property \(p_a\) during evaluation.
\end{definition}
Note that this definition does not limit tuples to properties of the constraint's context artifact but includes any artifact that is reached, respectively navigated to, during constraint evaluation.

From a quality and process point of view, we are primarily interested in ensuring the partial order of artifact changes, but rarely in the absolute time distance and delays between these changes, or the order of the underlying atomic events that caused each change. For example, we are interested in whether a review happens after every change, but not whether such a review is done within two days. As such, the temporal constraints we support are similar to the propositional fluents introduced by Giannakopoulou et al. in~\cite{DBLP:conf/sigsoft/GiannakopoulouM03}, in that they become true as a result of an action (i.e., artifact change), and remain true until a future action leads to them being evaluated to false.  Specifically, in our approach, an event describes a change to an artifact property.\footnote{For the remainder of this paper, we use the terms ``event'' and ``change'' interchangeably.} Therefore, we define a relevant change as:

\begin{definition}[Relevant Change]
    Let \(I\) be a set of incoming changes in the model. A change $i \in I$ is a \textit{relevant change} for tuple (\(a\),\(p_a\)) if it updates the value of property \(p_a\) of artifact \(a\).
\end{definition}

As a result, a constraint is only re-evaluated when the underlying artifact change affects an artifact in the scope of the constraint and possibly impacts the result of its evaluation. For example, a temporal constraint that only checks the order of a requirement's \texttt{state} property will not be triggered upon a change to the requirement's \texttt{priority} property.
Henceforth, whenever we refer to a \emph{change} in the context of a temporal constraint, we imply that this is a relevant change that possibly affects the outcome of some of the constraint's subexpressions. 
In light of the fluent properties of our constraints, we define the ``next moment in time'' below.

\begin{definition}[Next Moment in Time]
    Let $c \in C$ be a constraint, and the scope $S_c$ a set of tuples (\(a\),\(p_a\)), where $a \in A$ and $p_a \in P_a$ of constraint \(c\). The \textit{next moment in time} for constraint \(c\) is defined by the next set of changes which includes a relevant change for any of the tuples (\(a\),\(p_a\)) $\in$ $S_c$.
\end{definition}

Note especially that these changes cannot be compared to events used in discrete temporal constraint checking, as artifact changes are too low level to allow for meaningful constraints, mostly because they only describe the new property value but not the larger context. 
To further clarify this difference: a simple constraint pattern such as \texttt{next(A)} would have \texttt{A} substituted by an OCL expression such as \texttt{self.status='Ready'} in our case, while have \texttt{A} substituted by an explicit event such as \texttt{RequirementReady} in the discrete event process case. In the former case, we would interpret the constraint as \emph{the next (relevant) change needs to satisfy the OCL expression A, i.e., the requirement's status needs to be \texttt{Ready} next}. In the latter case, we would interpret the constraint as \emph{the next occurring event needs to be of type \texttt{RequirementReady}}.

This difference between the discrete and continuous temporal constraint checking also becomes apparent when considering the effect of subsequent changes, respectively events. In the continuous case, repeatedly setting a requirement's state to \texttt{Ready}\footnote{Without setting it to another value in between.} is treated just like a single change as the result of the OCL subexpression will remain unchanged (i.e., \texttt{true}), i.e., as with propositional fluents~\cite{DBLP:conf/sigsoft/GiannakopoulouM03}. In the discrete case, however, multiple \texttt{RequirementReady} events are treated separately. 
This has implications for mapping the discrete temporal patterns (of DECLARE and similar) as we will outline in Section~\ref{sec:rq1} to our application domain.

\subsection{Temporal Operators}

In order to add temporal semantics to OCL, we are reusing four basic operators and their semantics from LTL~\cite{DBLP:conf/sigsoft/GiannakopoulouM03} (i.e., \texttt{next}, \texttt{until}, \texttt{eventually}\footnote{We minimally redefine \texttt{eventually} as described further below.}, and \texttt{always})\footnote{In work on LTL \texttt{eventually} is also known as \texttt{finally} and \texttt{always} is also known as \texttt{globally}.} and define two new operators --- \texttt{\-atLeastOnce} and \texttt{everytime} --- as syntactic sugar over implications with \texttt{eventually} and \texttt{always}, respectively. 
The operators can be freely applied in an OCL constraint at any place where an OCL expression that defines a valid condition would be allowed. The valid parameters of our temporal OCL operators are in turn OCL expressions that can be evaluated to a boolean value (i.e., conditions). Nesting of temporal OCL expressions is thus possible.


\subsubsection{\texttt{next} Operator}
The \texttt{next} operator specified over a constraint A denotes that A shall hold at the next point in time, i.e., the immediately following constraint evaluation iteration. Typically, this operator by itself is rarely used. One possible use case is ensuring that the assignee of a bug, while not known upon bug reporting, is set upon any further changes to the bug. Listing ~\ref{lst:next} shows the use of the \texttt{next} operator for this purpose.  

\begin{lstlisting}[basicstyle=\footnotesize,caption={next(A) Operator} \label{lst:next} ]
context BugReport
inv: next(self.assignee.isDefined())
\end{lstlisting}

\subsubsection{\texttt{until} Operator}
The \texttt{until} Operator enables modeling a dependency between two OCL conditions, specifically that one condition only needs to hold until a second condition is fulfilled.
For example, a requirement must remain unreleased until it is reviewed first. 
As visible in Listing~\ref{lst:until}, The \texttt{until} operator requires that constraint A is true from the first evaluation onward and cannot become false before constraint B is fulfilled. This operator has the semantics of a ``strong until'' as specified in LTL, meaning that the overall constraint is not fulfilled until constraint B becomes true at least once. Note that we do not introduce an explicit ``weak until'' operator as its behavior can be obtained via composition of the basic operators: \texttt{always(A or until(A,B))}.

\begin{lstlisting}[basicstyle=\footnotesize,caption={until(A,B) Operator} \label{lst:until} ]
context Requirement
inv: until(not(self.status='Released'), 
                self.status='Reviewed')
\end{lstlisting}

\subsubsection{\texttt{eventually} Operator}
The \texttt{eventually} operator enables checking that an OCL condition is fulfilled at least once at some point in the future but not necessarily at the next point in time. Unlike the formal specification in~\cite{DBLP:conf/sigsoft/GiannakopoulouM03} which specifies $eventually(A) \models until(true, A)$, the semantics of \texttt{eventually} in our context are $eventually(A) \models until(false, A)$. This is especially useful in our application domain, as temporal constraints are often defined over multiple different artifact properties, or over collections of artifacts where the order of changes may be slightly different every time, or all artifacts cannot be changed at exactly the same time.
Continuing with our example from the motivating scenario, Listing~\ref{lst:eventually} encodes that we want to ensure that a requirement is eventually reviewed, regardless of how many different states it goes through earlier. 

\begin{lstlisting}[basicstyle=\footnotesize,caption={eventually(A) Operator} \label{lst:eventually} ]
context Requirement
inv: eventually(self.status='Reviewed')
\end{lstlisting}

\subsubsection{\texttt{always} Operator}
With the \texttt{always} operator, we express that the constraint, over which the operator is specified, needs to hold forever, not just once. 
A simple example use of this operator is the constraint that a bug report must be assigned to someone at all times during its lifetime (c.f. Listing~\ref{lst:always}). 
The difference to a standard OCL constraint checking for a bug report having an assignee is that without the \texttt{always} operator, we can only check that currently someone is assigned but not whether this is true throughout the bug's life-cycle. 

\begin{lstlisting}[basicstyle=\footnotesize,caption={always(A) Operator} \label{lst:always} ]
context BugReport
inv: always(self.assignee.isDefined())
\end{lstlisting}

\subsubsection{Syntactic Sugar - \texttt{atLeastOnce} and \texttt{everytime} Operators}
In the standard LTL notation, the \texttt{always} and \texttt{eventually} operators are overloaded to introduce a conditional evaluation of the actual constraint using the implies operator, e.g., to define that a condition B only needs to hold if condition A holds earlier (i.e., a trigger). The semantics of these operators are as follows:

$atLeastOnce(A,B) \models eventually(A \implies B \: or \: next(B))$

$everytime(A,B) \models always(A \implies B \: or \: next(B))$

For example, we want to ensure that a requirement that is accepted for implementation is also later reviewed. 
The \texttt{until} operator is not a suitable option here as it forces condition B to become true at some time, however, here we want to be able to express that condition A might never become true, and hence condition B might never need to become true either. 

Depending on whether condition B needs to hold just once after condition A holds or every time we apply the \texttt{atLeastOnce} or the \texttt{everytime} operator (c.f., Listing~\ref{lst:atleastonce} and Listing~\ref{lst:everytime}, respectively).  

\begin{lstlisting}[basicstyle=\footnotesize,caption={atLeastOnce(A,B) Operator} \label{lst:atleastonce} ]
context Requirement
inv: atLeastOnce(self.status='Accepted', 
              self.status='Reviewed')
\end{lstlisting}

We decided to add this syntactic sugar as we believe it makes reading LTL constraints easier as it becomes immediately clear that a trigger condition is present.

The \texttt{atLeastOnce} operator specifies that constraint B must hold immediately or next after the trigger is fulfilled, at least once. If constraint B is not fulfilled then, the operator waits for the next time the trigger condition is fulfilled (i.e., when condition A becomes false and true again). As soon as constraint B is fulfilled at least once after the trigger, the operator will not trigger again. In our example in Listing~\ref{lst:atleastonce}, when the artifact status is set to ``Draft'', the constraint checks whether the next status change is ``Ready for Review''. If it is not, the mechanism will trigger again when the ``Draft'' status is reached the next time, as long as the constraint is not fulfilled at least once.

If the intended behavior is that constraint B needs to be fulfilled every time the trigger is reached, then one would use the \texttt{everytime} operator. This operator expects that every time the trigger is fulfilled, the constraint will be fulfilled immediately or in the next evaluation. In the example in Listing~\ref{lst:everytime}, the expected behavior is that every time the artifact reaches the ``Draft'' status, then the status must become ``Ready for Review'' immediately thereafter.

\begin{lstlisting}[basicstyle=\footnotesize,caption={everytime(A,B) Operator} \label{lst:everytime} ]
context Requirement
inv: everytime(self.status='Draft', 
               self.status='Ready for Review')
\end{lstlisting}

\subsubsection{Operator nesting}
These six operators provide the basis for defining more complex constraints through nesting.
For example, the \texttt{atLeastOnce} and \texttt{everytime} operators require that constraint B is fulfilled immediately or in the next evaluation after the trigger A is fulfilled. This might be too restrictive.
For our example, this implied that every time the status of a requirement is ``Draft'', it cannot reach an intermediary status before it reaches ``Ready for Review''. 
When this is deemed too restrictive, nesting constraint B in an \texttt{eventually} operator results in an expression that requires that 'Ready for Review' is, at some point in the future,  at least once reached after every time the artifact reaches 'Draft' (c.f., Listing~\ref{lst:nesting} second example). Note that even though the \texttt{everytime} operator is used, it is not required that constraint B is fulfilled before condition A becomes false and true again. If this is desired more complicated patterns such as the one we discuss in Section~\ref{sec:rq1} need to be applied.
Similarly, using the \texttt{atLeastOnce} operator, once constraint B is fulfilled, constraint A may become true and false infinitely often without constraint B having to remain true, nor become true again.  In Listing~\ref{lst:nesting} (first example) this means that if a requirement was ``Accepted'' and then ``Reviewed'', it might, for example, become ``Rejected'', ``Draft'', ``Ready for review'', and ``Accepted'' again, but it does not need to reach status ``Reviewed'' again.

\begin{lstlisting}[basicstyle=\footnotesize,caption={Nesting of Operators} \label{lst:nesting} ]
context Requirement:
inv: atLeastOnce(self.status='Accepted', 
              always(self.status='Reviewed'))
context Requirement:
inv: everytime(self.status='Draft', 
        eventually(self.status='Ready for Review'))
\end{lstlisting}

\subsection{Constraint Execution Architecture}

In order to evaluate the truth value of the constraints specified using this temporal OCL extension, we are building on an experimental constraint checker proposed by Reder et al. \cite{DBLP:conf/models/RederE12}. The authors have given us permission to extend their solution with the temporal operators described above, as well as provided a copy of their prototype source code and grammar definition. We have selected this checker for its iterative checking mechanism, i.e., it already analyses which change potentially affects which constraint instance and automatically triggers a re-evaluation of those constraint instances. It is, however, completely unaware of temporal constraints. In the next subsection, we briefly introduce the basics of this checker and thereafter outline how we extended it to support temporal constraints.

\subsubsection{Preexisting Incremental Checking}
The ability to incrementally check constraints, i.e., to only check those constraints that are affected by a change, relies on two phases: first, tracking which artifact is used (anywhere) in the scope of which constraint instance, and second, looking up which artifact a change belongs to and obtaining the constraint instance (c.f. Definition~\ref{def:scope}).

In support of the former, when an artifact is created that matches the context type of a constraint (e.g., a requirement), the checker will then create a new constraint instance and a mapping from the artifact to the newly created constraints. As it then navigates across artifact properties (as determined by the constraint) to evaluate the constraint outcome, it creates for each additional artifact that it visits also a mapping to the constraint and marks the property as relevant to this constraint. The result is a map of artifacts to constraint instances. This mapping is updated upon artifact changes. 

In support of the latter, when an artifact change (or set of changes) occurs, the checker (a) checks if the change property is relevant to a constraint (as indicated by the property markings) and then (b) retrieves all constraint instances according to the mapping and (c) triggers the evaluation of those constraints. Note that a change might lead to a re-evaluation of multiple constraints, and multiple changes can lead to the evaluation of a single constraint. Each constraint, however, is only evaluated only once per artifact change set. Evaluation occurs then upon a view of the complete set of artifacts that already has all the changes applied.

Note that while we obtain for free the incremental capabilities of re-evaluating only those constraints that are potentially affected, extending the existing constraint checker with temporal operators that work in an incremental manner is non-trivial.

\subsubsection{Integrating Temporal Constraint Checking Capabilities}
The OCL constraints supported by the constraint checker are defined along a grammar expressed using yacc and flex, which allowed us to extend the language by adding our temporal operators.  
The constraint checker itself is implemented in Java where we subsequently provided additional Java classes that implement our six temporal operators. 

The challenge is to enable specifying temporal constraints in a manner that they can be seamlessly interwoven with regular OCL expressions and evaluated just like regular constraints. Namely, the constraint checking mechanism needs to be able to generate a hierarchical \textit{evaluation tree} which contains an evaluation node with the result of every OCL (sub)expression in the evaluated rule. 

This is non-trivial as, by default, OCL expressions only return Boolean truth values. These are sufficient for regular constraints but insufficient for temporal operators. 
Within temporal constraints, we need to be able to distinguish between \textbf{temporary} and \textbf{permanent} truth values. 
We consider the current boolean expression result as a \textbf{temporary} truth value when there exists at least one possible future change that can lead to a change of this truth value. Inversely, we consider the current boolean expression result in a \textbf{permanent} truth value if there exists no possible future change that can change the expression outcome.

   Let us consider the difference for an example requirement in status ``Ready for Review'' and two temporal constraints C1 and C2: \texttt{eventually(self.status = 'Reviewed')} and \texttt{always(self.\-status='Ready for Review')}. We consider C1 to be temporarily false, as C1 is currently false but the requirement might still reach the status ``Reviewed'' in the future. Similarly, we consider C2 to be temporarily true as C2 is currently fulfilled but there might be a future change that will violate it. Suppose the requirement then reaches status ``Reviewed''. C1 is now permanently true as there is no future change available that would change the evaluation result. Likewise, C2 is now permanently false as there is no future change possible that would fulfill the constraint. Hence, as soon as a temporal subexpression within an OCL constraint reaches a permanent truth value, that subexpression is no longer evaluated upon future changes. We denote this subexpression as being ``terminated'' (in comparison to being ``not terminated'' as long as it holds a temporal truth value).\footnote{Note that these example constraints ignore the possibility that a requirement reaches a different state again in the future. This is due to demonstration purposes in the paper and a more refined constraint is necessary to be useful in practice.}

The distinction between \textbf{temporary} and \textbf{permanent} truth values does not change the semantics of the operator. It is only available and used within a temporal constraint to understand the need for re-evaluation. Any outer expression, e.g., a ``forAll'', remains unaware of their child expression(s) being of temporal nature and hence will make no distinction, subsequently treating \textbf{temporary} and \textbf{permanent} truth values simply as true or false.

For \texttt{atLeastOnce} and \texttt{everytime} operators, we track whether the trigger condition A has been fulfilled and then start checking whether the conditional constraint B is true as well. We need to do this as the trigger condition A is not required to remain true until constraint B is true. Constraint B might be temporarily true or false, but we need to wait for it to become permanently true or permanently false. A trigger value of false indicates that the operator waits for the trigger condition A to become fulfilled (again).

\subsubsection{Operator Evaluation Result Truth Table}
As we support nesting of temporal constraints, our temporal operators need to react differently depending on whether their constraints A or B return a temporary or permanent evaluation result.\footnote{Note, that the evaluation result of a regular OCL constraint will be treated as a permanent true or false.} 
To this end, the truth table in Table~\ref{tab:truthtable} describes how the various combinations of A and B yielding temporary (i.e., temporarily true/false) and permanent (i.e., true/false) truth values determine the operators' temporary or permanent evaluation result in return. A ``?'' signifies that the truth value is irrelevant.
The table only describes the outcome as long as an operator is not terminated yet. Upon operator termination with permanently true or permanently false, the evaluation results for A or B no longer matter as the operator's result will not change anymore.

Specifically for \texttt{atLeastOnce} and \texttt{everytime}, the table shows that these two operators return the temporary truth value of constraint B as long as B is not terminated. The table also shows that as long as the trigger condition A has not been fulfilled yet, respectively again, the operator will return temporarily true. Furthermore, it is irrelevant whether trigger condition A returns a temporary or permanent truth value.

Table~\ref{tab:truthtable} also highlights that \texttt{eventually} and \texttt{atLeastOnce} can never return permanently false, and that \texttt{always} and \texttt{everytime} can never return permanently true.

\begin{table}[t]
    \centering

    \caption{Temporal operator result (Op. result) truth table for different truth values of A and B: permanently true (permT), permanently false (permF), temporarily true (tempT), and temporarily false (tempF). T indicates permanently or temporarily true, F indicates permanently or temporarily false.
    ``?'' indicates the truth value is irrelevant. $\dagger$: and when trigger condition A has never been (temporarily) true before. }
    \label{tab:truthtable}
    \addtolength{\tabcolsep}{-4pt}
    \begin{tabular}{c | c | c | c | c}
        Operator & Prev. result A  & Curr. result A & result B & Op. result \\
        \hline
        next(A) & n/a & permT & n/a & permT \\
         & n/a & permF & n/a & permT \\
          & n/a & tempT & n/a & tempT \\
          & n/a & tempF & n/a & tempF \\        
        \hline 
        eventually(A) & ? & permT & n/a & permT \\
          & ? & tempT & n/a & tempT \\
          & ? & F & n/a & tempF \\        
        \hline
        always(A) &  T & T & n/a & tempT \\
          &  T & F & n/a & permF \\
        \hline 
        until(A, B) & ? & T & F & tempF \\
         & ? & F & F & permF \\
         & ? & ? & tempT & tempT \\
         & ? & ? & permT & permT \\
        \hline 
        atLeastOnce(A, B) &  F $\dagger$ &  F & ? & tempT \\
        &  F &  T & permT & permT \\
         &  F &  T & tempT & tempT \\
         &  F &  T & F & tempT \\
         &  T & ? & permT & permT \\
        &  T & ? & tempT & tempT \\
         &  T & ? & F & tempF \\

        \hline
        everytime(A, B) &  F $\dagger$ &  F & ? & tempT \\
         &  F &  T & T & tempT \\
         &  F &  T & F & tempF \\
         &  T & ? & T & tempT \\
         &  T & ? & permF & permF \\
         &  T & ? & tempF & tempF \\

        \hline
    \end{tabular}
\end{table}

\subsubsection{Maintaining operator state} \label{sec:opstate}
Table~\ref{tab:truthtable} makes apparent that for operators \texttt{always}, \texttt{atLeastOnce}, and \texttt{everytime}, we need to keep the state of the previous evaluation results of subexpression A. The underlying constraint checker does not support this as for regular OCL constraint evaluation only the current evaluation result is of relevance.
Hence, for these three operators, we store the latest evaluation result of constraint A. For operators \texttt{atLeastOnce} and \texttt{everytime} we also need to store whether the trigger condition A has been fulfilled before as it may evaluate already to false again before constraint B is permanently true (or false).
Note that we need not maintain the history of changes not the overall history of constraint evaluations. This makes our approach very memory efficient as the state maintenance overhead is negligible (i.e., a single Boolean value per \texttt{always} operator instance, respectively two Boolean values per \texttt{atLeastOnce} and \texttt{everytime} operator instances). As a consequence, memory usage merely grows linearly with the number of operator instances.

\section{Evaluation}
\label{evaluation}
\subsection{Research Questions}
We assess our mechanism along the following two research questions that address expressiveness and performance, respectively.

\textbf{RQ1 - Are our temporal operators able to express a variety of higher level, realistic temporal constraints? } We are interested in understanding whether our small set of fundamental temporal operators may be composed to express more complicated constraints that are also practically relevant. 
By analyzing whether we can use our operators to obtain the semantics of the temporal patterns available in DECLARE ~\cite{declare}, we determine whether our approach would leave significant practical scenarios uncovered.

\textbf{RQ2 - Is our prototype sufficiently quickly evaluating realistic temporal constraints to enable timely feedback to engineers?} 
We are interested in understanding how long temporal constraint checking takes to estimate our approach's practical applicability. To this end, we replay real, historical engineering artifact changes and measure the average constraint evaluation duration.

\subsection{Evaluation Design}

For RQ1, we are evaluating the expressiveness of our approach by comparing it with the 18 temporal constraints made available by DECLARE~\cite{declare,de2014reasoning}, a state-of-the-art declarative process modeling and execution approach.
DECLARE defines a comprehensive set of common patterns useful for specifying temporal constraints over process step sequences. These patterns are an extension of the initial patterns by Dwyer et al. which in turn are found to cover 92\% of 555 temporal specifications collected from scientific literature, verification tooling, and educational settings~\cite{dwyer1998property}.  
Bianculli et al.~\cite{6227125} collected a larger set of patterns (including those by Dwyer et al.). However, their collected patterns beyond Dwyer et al. focus primarily on the amount of time between events or the metrics over the number of event occurrences, which are rarely relevant patterns in the context of engineering processes.

The DECLARE patterns are grounded in the same LTL operators that we define (a definition of a comprehensive set of Declare patterns in LTL formulae can be found in De Giacomo et al.\cite{de2014reasoning}). However, the patterns and subsequent LTL formulae focus on application to traces of \textbf{discrete} events, and not to OCL constraints that hold over \textbf{continuous} periods of time. We, therefore, modified the LTL formulae to achieve a most similar behavior and discuss why this was necessary.
Ultimately, we determine the answer to RQ1 by counting the number of DECLARE patterns that we can support.

For RQ2, we evaluate the runtime performance for checking a set of temporal constraints against a series of engineering artifact changes. Specifically, we track how often a particular temporal constraint is invoked over multiple artifact changes. We then measure how much time on average a temporal constraint needs for evaluation and how long overall the constraint checking for our set of changes takes.

To obtain insights into realistic behavior, we collected a set of real temporal constraints and artifact changes from the engineering process of our industry partner ACME-RA. The temporal constraints were described in natural language which we subsequently translated into OCL and LTL.
ACME-RA\footnote{The company's identity and data have to remain confidential due to the sensitive nature of the analyzed data.} hosts a web platform for recreational activities. They track the development and management of their platform using Jira issues, while following an agile development process. In this process, each issue is created with the status set to 'Open'. The development of the issue is marked by the status 'In Development', after which the status is set to 'Ready For Review'. Once the review is completed, the issue is set to the status 'Reviewed', and then the testing phase starts with the issue moving to status 'In Testing'. After this, the issue is 'Resolved', and then can be 'Closed' once all sub-issues are also closed. Additionally, the review and testing steps may be skipped. Based on the findings in these steps, the issue could be 'Reopened' or return to any previous status with the additional 'again' keyword to mark revision, such as 'In Development Again'. Finally, the process can be suspended during the development or testing phase ('Suspended Development' or 'Suspended Test').
We mapped these Jira issue states to process steps and added in total 21 temporal pre and post conditions, as well as quality assurance conditions (see Supporting Online Material \cite{SOM}).  Consequently, the conditions ranged from simple temporal constraints, such as the initial precondition \textit{eventually(self.status = 'Open')}, to complex compositions of temporal and regular OCL operators. 
Finally, an engineer at ACME-RA confirmed that the defined constraints reflect the rules that developers at ACME-RA are supposed to follow.

The set of engineering artifacts consists of a JSON dump of four multi-year ACME-RA projects, ranging from 939 to 2676 Jira issues out of which we focus on issues of type “Task”, that were successfully resolved (i.e., in state “Fixed”), with a non-empty set of child issues. This resulted in 46, 21, 119, and 81 process instances, respectively.

We then reset each artifact to its initial state and applied each artifact change in the correct temporal order to mimic how the individual engineering processes actually happened. Upon each individual change, we invoked our constraint checker prototype and determined constraint fulfillment and hence process progress.

\subsection{RQ1 Results - Expressiveness} \label{sec:rq1}
A selection of the results of adapting the DECLARE patterns to our approach are provided in Table~\ref{tab:declare}  where we list for every pattern the corresponding LTL formulae using our operators, an explanation of the pattern, an example along the lines of our motivating scenario, and subsequent OCL expressions for placeholders A and B. We also provide the original LTL formulae where we needed to adapt them for application in our continuous (rather than discrete) constraint checking context. The adaptation of all the 18 DECLARE patterns with additional formatting for readability is available as SOM~\cite{SOM}.

The primary reason behind our modifications is that an event may happen multiple times, while a constraint remains true and must become false before becoming true again. For example, the \emph{Absence 2} pattern is violated as soon as event A happens a second time. In our continuous setting and without modification, this pattern would be violated upon the first reevaluation after constraint A becomes true as upon this reevaluation, A might still be true, and hence the second eventually becomes fulfilled. For this pattern, we refined its semantics to ensure that once constraint A is true, it remains true or, if false again, stays false. Note that to express that constraint A may not become false again, the latter ``or'' part simply needs to be removed.

Similarly, modifications were necessary for the pattern group \emph{Alternating Response}, \emph{Alternating Precedence}, and \emph{Alternating Succession}. These patterns enforce that events A and/or B are not allowed to come in succession without the other event interleaving. 
When A and B are constraints, we run into the same problem as above: constraint A might remain fulfilled, subsequently violating the until operator. Hence we modify the pattern to allow A to remain true or become false, but B needs to become true before A can become true again. 

We further modified the \emph{chain} patterns, including the \emph{Negation Succession}, which address constraints on how two events A and B must or must not happen right after each other. While we can express that conditions A and B become true right after each other, in our case, evaluation points in time are determined by any change to a property. Hence, it is likely that a non-trivial constraint experiences interleaving changes that violate the constraint but were not the intention of the constraint designer. A more practical pattern interpretation in our case is to enforce that constraint A becoming true must (or must not) be followed by constraint B coming true without A becoming false in the meantime. This interpretation is robust to changes that do not affect A or B but perhaps some other part of a more complicated constraint.  

Overall the modified patterns retain the semantics of the original patterns but enable application in the \emph{continuous} setting of OCL constraints (rather than discrete events).

\vspace{0.2cm}
\observe{
        \textbf{RQ1 Summary:} All 18 DECLARE patterns can be expressed with our operators.}

\hvFloat[floatPos=htbp,rotAngle=90,capPos=top,capWidth=w]%
{table}%
{
\centering
\scriptsize
\begin{tabular}{p{4.5em}p{20.0em}p{24em}p{27.0em}p{13.7em}}
\textbf{Pattern} & \textbf{Constraint} & \textbf{Explanation} & \textbf{Example} & \multicolumn{1}{l}{\textbf{Example Constraint}} \\ 
    \hline
    Existence & eventually(A) & Condition A becomes true at least once. & At some point in time, the requirement will be in status ``Reviewed''. & \multicolumn{1}{l}{A: self.status='Reviewed'} \\ 
    \hline
    Absence 2 & atLeastOnce(A, always(A) or \newline{} not (atLeastOnce (not(A), A))) \newline{} \emph{Orig: not eventually(A and eventually(A)))} & Condition A stays true or, when becoming false, does not become true again. & A requirement can never be released more than once but may be retired. & \multicolumn{1}{l}{A: self.status='Released'} \\ 
    \hline
    Alternating precedence & until(not(B), A) and everytime(B, always(B) or until(not(B), A)) \newline{} \emph{Orig: until(not(B), A) and everytime(B, next(until(not(B), A)))} & Condition A must always become true before condition B becomes true and B cannot become false and true again until A has not become false and true again.  & User documentation can only be released when reviewed before and needs such a review before every re-release. & A: self.status='Reviewed' \newline{}B: self.status='Released' \\ 
    \hline
    Alternating succession & everytime(A, eventually(B) and (always(A) or until(not(A), B))) and until(not(B), A) and \newline{}everytime(B, always(B) or until(not(B), A)) \newline{} \emph{Orig: everytime(A, next( until(not(A), B))) and until(not(B), A) and everytime(B, next(until(not(B), A)))}  & Condition A and B alternate their truth values (combined Alt. response and Alt. precedence). & A requirement needs to be reviewed after a change, and will only be reviewed if there was a change before, and cannot be re-edited without a review before. & A: self.status='Draft' \newline{}B: self.status='Reviewed' \\ 
    \hline
    Chain response & until(not(B), A) and everytime(A, until(A,B)) \newline{} \emph{Orig: everytime(A, next(B)) }& Condition A becoming true must be followed by condition B becoming true without any other values inbetween. & A requirement in state ``Draft'' must be next in state ``Ready for Review'' before reaching any other states. & A: self.status='Draft' \newline{}B: self.status='Ready for Review' \\ 
    \hline
    Chain precedence & everytime(A, always(A or eventually(B)) or not(eventually(B))) and until(not(B), A) \newline{} \emph{Orig: everytime(next(B), A)} & Condition B can only become true in the next iteration after condition A becomes true. & A requirement can be released only when it was previously in status ``Reviewed''. & A: self.status='Reviewed' \newline{}B: self.status='Released' \\ 
    \hline
    Chain succession & everytime(A, until(A,B)) and \newline{} everytime(B, until(B,A)) \newline{}\emph{Orig: everytime(A, next(B)) }& Condition A becoming true must be followed by condition B becoming true without becoming false in the meantime and vice versa. & Whenever a requirement is approved, it also needs to be released next and can only be released when approved just before. & A: self.status='Approved' \newline{}B: self.status='Released' \\ 
    \hline
    Negation succession  & atLeastOnce(A, not(eventually(B))) \newline{}\emph{Orig: everytime(A, not(eventually(B)))} & Once condition A becomes true, it can never be followed by condition B becoming true. & A requirement may initially be temporarily rejected, but as soon as it is released, it can never be rejected thereafter. & A: self.status='Released' \newline{}B: self.status='Rejected' \\ 
    \hline
    Negation chain succession & everytime(A, not(until(A,B))) and \newline{} everytime(B, not(until(A,B))) \newline{}\emph{Orig: everytime(A, next(not(B))) }& Every time condition A becomes true, it can never be followed by condition B becoming true and vice versa. & A requirement cannot transition directly from ``Ready for Review'' to ``Released'' and vice versa. & A: self.status='Ready for Review' \newline{}B: self.status='Released' \\
\end{tabular}%
}%
{A selection of DECLARE patterns, their LTL representation (and original version), pattern explanation, example, and example constraints. All 18 DECLARE patterns in the same format are available in the SOM~\cite{SOM} .}%
{tab:declare}

\subsection{RQ2 - Performance}

During the replay of the artifact changes in the four projects provided by ACME-RA, we have measured the runtime performance of our mechanism. 
The result of this evaluation, presented in Table~\ref{tab:eval}, consists of the aggregated results of 10 replays for each of the four ACME-RA projects on a standard desktop computer with an Intel Core i7-4770 CPU with 4 Cores and 16GB RAM. The replay of changes for a project took on average between 5 and 16.5 seconds. Out of the $\sim$8,000 to $\sim$29,000 replayed events per project, between $\sim$850 to $\sim$5,300 events resulted in at least one temporal constraint being evaluated. Between $\sim$4,500 to $\sim$30,500 constraint instances were evaluated per project and replay iteration. Each of these temporal constraint evaluations took, on average, below 0.2 milliseconds. 

Recall that memory overhead per operator instance is minimal (c.f. Sec~\ref{sec:opstate}) as our approach incrementally checks the constraint result on the fly without having to analyze the whole change history every time and with only minimal state keeping in memory. Hence we do not report on memory usage as it would yield little insight on performance. 

We neither compared against a non-incremental baseline as outlined in the Section~\ref{sec:sota} on related work , no existing temporal constraint checker enables expressing such navigation rich constraints as needed in our case study. 

\begin{table}[H]
    \centering
    \caption{Performance evaluation via replaying of artifact changes, detailing total replayed events ($\#$total), number of events that triggered a least one constraint ($\#$triggers), number of constraint evaluations ($\#$eval), average evaluation time across all constraints per event in milliseconds (avg(ms)), and average total time replaying all events in seconds (total(s)).}
    \label{tab:eval}
    \begin{tabular}{c | r | r | r | r | r}
    Project & $\#$total & $\#$triggers & $\#$eval & avg(ms) & total(s) \\
    \hline
    P1 & 8,394 & 1,390 & 6,333 & 0.33 & 5.8 \\
    \hline
    P2 & 9,362 & 868 & 4,532 & 0.34 & 5.3 \\
    \hline
    P3 & 28,623 & 5,342 & 30,580 & 0.19 & 16.5 \\
    \hline
    P4 & 11,225 & 4,214 & 20,390 & 0.20 & 9.3 \\
    \hline
    \end{tabular}
\end{table}

\vspace{0.2cm}
\observe{
        \textbf{RQ2 Summary:} Incremental temporal constraint checking is feasible and fast, requiring always less than 0.4 milliseconds per constraint evaluation.}

\subsection{Discussion}
The results show that our approach can express high-level meaningful temporal patterns and evaluate temporal constraints sufficiently fast for application in real-world settings. 
This work hence lays the foundation for more sophisticated engineering process support. At this initial level of our research line, the impact on practitioners is still small as we have (necessarily) left a major factor unaddressed in this paper: previous research has shown that LTL formulae are hard to understand and write \cite{czepa2018understandability}. Supporting the efficient design of temporal constraints, ensuring that they reflect what the designer intended to encode, and conveying that meaning to engineers that should abide by these constraints (respectively need to understand the extent to which they deviated) requires significant additional research efforts. 
Such efforts are, however, outside the scope of this paper that focuses on demonstrating the feasibility of our approach. 

Additionally, we have extensively tested the temporal constraint specifications in a variety of scenarios extracted from the four industrial projects. Through this, we have found the operators work as intended.

The implications for researchers, however, are manifold: we provide a temporal constraint checker that may serve as the basis for research on repairing temporal constraints, i.e., ultimately supporting engineers in how to return to a consistent process state. Our approach also has the potential to be applicable for incremental run-time system checking, e.g., to ensure systems evolve or adapt correctly. Our approach addresses a shortcoming of existing approaches which only work on event traces without access to the overall system scope. It remains to be seen to what extent our operators are sufficient in such a different context as we left support for absolute time and time differences aside.

\vspace{0.2cm}
\textbf{Threats to validity:} 
We address researcher bias by modeling processes and constraints utilized in the industry. Especially for the performance evaluation, we obtained  feedback from our industry partner to ensure the constraints match those in the daily work of their engineers.

These constraints, however, might not reflect the needs and rules in place at other companies. 
Hence, the evaluation of other temporal constraints might not be as performant. However, our temporal operators make use of the regular OCL operators for deriving the underlying truth values. Also recall, that due to incremental checking, the amount and history of changes do not affect evaluation performance. Hence the performance is primarily determined by the complexity of any OCL rule used inside of a temporal operator, and the amount of artifacts,  that are part of a constraint's scope.

Further investigations with feedback from the industry will also reveal to what extent the DECLARE patterns are useful 
and which other patterns might be needed and to what extent these are supported by our basic LTL operators.

\section{Conclusions}
\label{conclusion}
Due to the iterative and flexible nature of engineering processes, engineers cannot be forced to strictly follow existing process models. However, in safety-critical system development, safety standards and regulations require that certain temporal constraints over the interrelated engineering artifacts are fulfilled throughout the process. Currently, these constraints are checked manually, resulting in a time-consuming task for the quality assurance personnel and a lack of timely feedback for the engineers. Existing approaches are either too restrictive, checking the process model but not allowing deviations, or cannot capture the data necessary for the navigation across artifact relations that is necessary for this context.

In this paper, we have proposed a mechanism for iterative, run time checking of temporal process constraints, based on an extension of OCL with LTL operators. We have evaluated our mechanism in terms of expressiveness, by supporting all the DECLARE~\cite{declare,de2014reasoning} patterns through our OCL extension. Further, our performance evaluation shows that a defined temporal constraint can be checked within $\sim$0.2 milliseconds. 

As future work, we will investigate mechanisms to support writing and understanding constraints, especially focusing on heuristics for deriving these temporal constraints semi-automatically from process definitions. 

\section*{Data Availability}
Supporting online material (SOM) \cite{SOM}  provides additional information on evaluation data, constraints, and DECLARE patterns.
The prototype, unfortunately, cannot be made public at the time of submission as it would jeopardize double-blind submission requirements but will be made available upon acceptance.

\begin{acks}
Anonymized
\end{acks}


\end{document}